# Determining the Impacts of Social Media on Students' Mood, Time Management and Academic Activities and the Relationship with their Academic Performance: The Nigerian Perspective

[1]Comfort Olebara
Department of Computer Science,
Faculty of Physical Science,
Imo State University, Owerri,
Imo State, Nigeria.
chy_prime@yahoo.com

[2]Obianuju Ezugwu
Department of Computer Science,
Faculty of Physical Science,
University of Nigeria, Nsukka,
Enugu State, Nigeria.
assumpta.ezugwu@unn.edu.ng

[3]Adaora Obayi
Department of Computer Science,
Faculty of Physical Science,
University of Nigeria, Nsukka,
Enugu State, Nigeria.
adaora.obayi@unn.edu.ng

[4]Deborah Ebem
Department of Computer Science,
Faculty of Physical Science,
University of Nigeria, Nsukka,
Enugu State, Nigeria.
deborah.ebem@unn.edu.ng

[5]Ujunwa Mgboh
Department of Computer Science,
Faculty of Physical Science,
University of Nigeria, Nsukka,
Enugu State, Nigeria.
obianuju.odike@unn.edu.ng

[6]Elochukwu Ukwandu
Department of Computer Science,
Cardiff School of Technologies,
Cardiff Metropolitan University,
Cardiff, Wales, United Kingdom.
Correspondent email:
eaukwandu@cardiffmet.ac.uk

*Abstract*— The twenty-first century can be said to be the time where digitisation is at its peak, with many programmers tapping into the desire of humans to remain connected without time and space limitations. The number of social media sites have increased exponentially, with new ones cashing in on the weaknesses of older ones, and others going beyond community guidelines by offering uncensored content. The vendors of these platforms, in order to have a wider reach do not place restrictions on viewing age, give promise of making young people famous, and other such attractive offers that make the youths addicted to the site. The possibility of hacking into accounts of users and using same for fraud is another rave among the Nigerian youths with desire for quick riches. The crash in prices of data, smart phones, and related digital devices has increased availability and access thereby closing digital divide and widening its adverse effects on the youths' morals and academic pursuits. It is important that the Nigerian government understand that factors that contribute to the dwindling performance level of students in government owned institutions so as to put in place policies and infrastructure that would help combat the challenges. This study investigated the effects of social media on students' academic activities, mood, and time management abilities. The result indicated that association between social media and academic activities is statistically significant, however, a negative association exists between them which implies that the high the level of social media activity, the lower academic activities participation. Similar association was observed on the effects of social media on students' time management ability. Statistical tests showed there is statistical significance, but also with a negative association which implies that increase in social media activities result in decrease in ability to effectively manage time. Result on the effect of social media on students' mood (angry, depressed) showed no statistical significance and a negative t-value, indicating a reversal on directionality of effects sought. This means that social media posts do not result in bad mood, but higher the level of bad mood leads to higher social media posts.

## I. INTRODUCTION

Surveys and studies have shown that social media is very attractive to the youthful population of many nations including Nigeria. The reason being that they are free once subscribed, easy to use and have high social appeal being social networking places according to Daniel *et al*. [1]. There are varieties of these social networking sites that enable users to create and share contents such as career information, marketing, education, ideas and other forms of expressions. No doubt, social media has become a very big influence on our world and majorly among the youthful population. Unfortunately, this youthful population form majorly student population as well and hence the need to know the influence these platforms have in them. Current studies on the impact of social media on our society have been focused on student academic performance using the Grade Point Average (GPA), some literature argue that it has negative effects on their academic performance according to Yeboah & Ewur [2], Kirschner & Karpinski [3], Rosen *et al.* [4], and Leyrer-Jackson & Wilson [5], while others argue that it rather aid them positively in their ability to multi-task depending on their level of education as in Junco [6], and Lau [7].

Two major issues appear prominent in all these studies: the amount of time spent on using social media and the likely effects it has on their time management ability. Secondly, the quality of information accessible through social media and the effects on students' behaviour as many of this information are uncensored. These factors will be discussed in detail as we work on finding their implications on Nigeria students' academic performance from when they become school leavers through their university education.

Section I introduces the background of the work. Section II represents the various literatures associated with the work while section III presents instruments and methods applied in order to answer the research questions. In section IV, the data presentation and analysis were conducted, hypotheses tested, and the result used to determine study outcome. Conclusion and recommendation were presented in Section V, concurrent or later production of electronic products, and (3) conformity of style throughout a conference proceedings. Margins, column widths, line spacing, and type styles are built-in; examples of the type styles are provided throughout this document and are identified in italic type, within parentheses, following the example. Some components, such as multi-leveled equations, graphics, and tables are not prescribed, although the various table text styles are provided. The formatter will need to create these components, incorporating the applicable criteria that follow.

## II. LITERATURE REVIEW

*1) Social media usage by students*

Surveys and studies have shown that social media is very attractive to the youthful population of many nations including Nigeria. The reason being that they are free once subscribed, easy to use and have high social appeal being social networking places [1]. There are varieties of these social networking sites that enable users to create and share contents such as career information, marketing, education, ideas and other forms of expressions. No doubt, social media has become a very big influence on our world and majorly among the youthful population. Unfortunately, this youthful population form majorly student population as well and hence the need to know the influence these platforms have on them. Current studies on the impact of social media on our society have been focused on student academic performance using the Grade Point Average (GPA), some literature argue that it has negative effects on their academic performance [2], [3], [4], and [5]), while others argue that it rather aid them positively in their ability to multi-task depending on their level of education as observed by [6], and [7].

Two major issues appear prominent in all these studies: the amount of time spent on using social media and the likely effects it has on their time management ability. Secondly, the quality of information accessible through social media and the effects on students' behaviour as many of this information are uncensored. These factors will be discussed in detail as we work on finding their implications on Nigeria students' academic performance from the point, they are school leavers through their university education.

*2) Time management and social media usage by students*

The ability to balance between leisure and study time so as to prepare well for examination known as time management has been judged as an important predictor of students' academic performance, [8]. The works of [9], [10] and [11] gave supportive evidence to the above information. While [9] agree that time spent correlates with student's academic performance, they went further to add that such time spent were heavily influenced by their attention span. [10] conducted their study on 72 students participating in a project known as SmartUnitn, and at the end underlined the need for control on students' access to smart phones, which gives them opportunity to go through social media platforms while studying as it showed negative effects on their time management ability and hence academic performance. The work of [11] was focused on 68 students of the University of Zululand, South Africa. Their findings showed that there is a relationship between time spent on academic activities with that of social media and classroom activities. They concluded from their results that time spent on social media can be used to predict students' academic performance.

A recent study by [12] suggests that the major influence of student's use of social media is socialisation. Scholars have argued that such activities impact negatively on them. Aside this, the contents available to the students through social media are hardly censored and how these affect their behaviours and by extension academic performance will be our focus in the next subheading.

## III. METHODOLOGY

### A. Research Purpose and Design

The purpose of this study is threefold. Firstly, to determine the effects of participants' social media activities on their participation in academic activities, secondly, to determine if participants' mood is affected after the read social media comments, and thirdly to determine the effects of social media involvement on students' time management abilities.

### B. Study Area

Students in two Nigerian universities were identified as participants for this research. University of Nigeria, Nsukka and Imo state university undergraduates, graduates and post graduate students participated in the study. One community based and the other a national university. This was done in order to increase the number of participants, considering that the study was carried out at a time Nigerian Academic Staff Union of Universities was on strike and access to target group was limited. Received responses although low, were used to carry out a pilot study in order to test the hypotheses provided for this study.

### C. Research Instrument and Method

Data for this study was collected using a questionnaire designed with google form and distributed to the participants' WhatsApp number and emails addresses through their class WhatsApp groups and the university ICT unit respectively. The questionnaire consists of five sections. Section1 was used to collect demographic information such as age, age smart phone was acquired, study level in the university, and so on. Section2 collected information on participants' time management factors such as number of social media accounts they operate, hours spent on social media daily, time of the day the access, and level of involvement on social media activities. In section 3, data on impact of social media activities on academic activities is collected by requesting information on multitasking social media with lectures, adoption of social media for academic problem solving. Sections 4 and 5 are used to collect data on impacts of social media on students' behavior and mood. Close-ended questions as well as Likert scale statements were used in generating the questionnaire.

### D. Sample Size

A total of 396 participants participated in the study. Responses received were subjected to statistical analysis. The expected sample size could not be obtained as a result of strike action of Nigerian universities academic staff.

### E. Research Questions

The following questions were raised in order to investigate the research hypotheses:
1. Do social media engagement affect the academic participation level of students?
2. Do social media posts affect student's mood (make them angry or depressed)?
3. Does social media engagement affect students' time management ability?

## F. Hypotheses

H1: Social media activities affect the academic activities of the students.

H1: Social media posts affect student's mood (make them angry or depressed)

H1: Social media involvement level affect students' time management ability.

## G. Data Analysis

Data from participants were coded into numerical data in a spreadsheet. Processing and analysis of captured data was carried out using SPSS (Statistical Package for Social Sciences) version 20.0 and reported by means of descriptive statistics, frequency distribution, regression analysis and independent t-test.

## IV. RESULT AND DISCUSSION

In this section, the output from the data analysis will be discussed.

Tables 1-7 below present the demographic characteristics of participants.

*Table 1: Gender*

|  |  | Freq. | Percent | Valid Percent | Cumulative Percent |
|---|---|---|---|---|---|
| Valid | Female | 226 | 57.1 | 57.2 | 57.2 |
|  | Male | 169 | 42.7 | 42.8 | 100.0 |
|  | Total | 395 | 99.7 | 100.0 |  |
| Missing | System | 1 | .3 |  |  |
| Total |  | 396 | 100.0 |  |  |

*Table 2: Age*

|  |  | Freq | Percent | Valid Percent | Cumulative Percent |
|---|---|---|---|---|---|
| Valid | 16-18years | 57 | 14.4 | 14.7 | 14.7 |
|  | 19-21years | 149 | 37.6 | 38.4 | 53.1 |
|  | 22-25 | 163 | 41.2 | 42.0 | 95.1 |
|  | 25years+ | 19 | 4.8 | 4.9 | 100.0 |
|  | Total | 388 | 98.0 | 100.0 |  |
| Missing | System | 8 | 2.0 |  |  |
| Total |  | 396 | 100.0 |  |  |

Table 2 shows that the ages of participants were divided into five groups with the following distributions: 16-18years has a distribution of 57 out of 396 participants, representing 14.4%. Group 2 captured those between the ages of 19-21years with a distribution of 149 participants, representing 37.6%. In group 3, participants between 22-25year were captured. A total of 163 participants participated in this study representing 41.2% of participants. The last group captured data from participants above 25years. 19 responses were received from this group representing 4.8% of total participants.

*Table 3: Academic attainment*

|  |  | Freq. | Percent | Valid Percent | Cumulative Percent |
|---|---|---|---|---|---|
| Valid | Undergraduate | 370 | 93.4 | 96.6 | 96.6 |
|  | Graduate | 10 | 2.5 | 2.6 | 99.2 |
|  | Postgraduate | 3 | .8 | .8 | 100.0 |
|  | Total | 383 | 96.7 | 100.0 |  |
| Missing | System | 13 | 3.3 |  |  |
| Total |  | 396 | 100.0 |  |  |

The academic attainment of the participants has the following distributions: 93.4% (370) are undergraduates, 2.5% (10) are graduates, and 0.8% (3) are postgraduates. Data on the age respondents acquired smartphone was collected using two groups. Participants that acquired smartphone at ages below 18years and those that acquired smartphone when they were 18years and above. 67.9% (269) acquired when below 18 and 31.3% (120) acquired when they 18years and above.

*Table 4: Age of acquiring a mobile phone*

|  |  | Freq. | Percent | Valid Percent | Cumulative Percent |
|---|---|---|---|---|---|
| Valid | Below 18years | 269 | 67.9 | 68.3 | 68.3 |
|  | 18-25yrs | 124 | 31.3 | 31.5 | 99.7 |
|  | 25 years + | 1 | .3 | .3 | 100.0 |
|  | Total | 394 | 99.5 | 100.0 |  |
| Missing | System | 2 | .5 |  |  |
| Total |  | 396 | 100.0 |  |  |

*Table 5: Case Study Higher institutions*

|  |  | Freq. | Percent | Valid Percent | Cumulative Percent |
|---|---|---|---|---|---|
| Valid | UNN | 43 | 10.9 | 10.9 | 10.9 |
|  | IMSU | 350 | 88.4 | 89.1 | 100.0 |
|  | Total | 393 | 99.2 | 100.0 |  |
| Missing | System | 3 | .8 |  |  |
| Total |  | 396 | 100.0 |  |  |

On the Universities used, UNN had a frequency of 43, representing 10.9% of participants while IMSU had a frequency of 350 representing 88.4% of respondents.

*Table 6: Hours spent on Social Media per day*

|  |  | Freq. | Percent | Valid Percent | Cumulative Percent |
|---|---|---|---|---|---|
| Valid | 0-3hrs | 164 | 41.4 | 42.1 | 42.1 |
|  | 4-8hrs | 148 | 37.4 | 37.9 | 80.0 |
|  | 9-12hrs | 49 | 12.4 | 12.6 | 92.6 |
|  | 13hrs + | 29 | 7.3 | 7.4 | 100.0 |
|  | Total | 390 | 98.5 | 100.0 |  |
| Missing | System | 6 | 1.5 |  |  |
| Total |  | 396 | 100.0 |  |  |

With respect to number of hours spent on social media daily, 41.4% (164) indicated 0-3hours on social media daily, 37.4% (148) spend 4-8hours on social media, 12.4% (49) spend 9-12 hour daily, while 7.3%(29)spend 13hours and above on social media daily.

*Table 7: Total number of social media account used on regular basis*

|  |  | Freq. | Percent | Valid Percent | Cumulative Percent |
|---|---|---|---|---|---|
| Valid | 0 | 1 | .3 | .3 | .3 |
|  | 1 | 166 | 41.9 | 1.9 | 42.2 |
|  | 2 | 89 | 22.5 | 22.5 | 64.6 |
|  | 3 | 66 | 16.7 | 16.7 | 81.3 |
|  | 4 | 50 | 12.6 | 12.6 | 93.9 |
|  | 5 | 13 | 3.3 | 3.3 | 97.2 |
|  | 6 | 10 | 2.5 | 2.5 | 99.7 |
|  | 7 | 1 | .3 | .3 | 100.0 |
|  | Total | 396 | 100.0 | 100.0 |  |

Demographic information on number of social media accounts regularly used by participants was collected. 41.9% (166) use one account, 22.5% (89) use two accounts, 16.7% (66) use three accounts, 12.6% (50) use four accounts, 3.3% (13) use five accounts, 2.5% (10) use six accounts and 0.3 (1) use seven social media accounts.

*1) Testing of Hypothesis*

**1. H1: Social media engagements affect academic activities.**

To test for this, first we added up the scores that make up the total social media scores (TSM), also the total scores for the Academic activities (TAA) was computed. In order to get the scores to their original form, the sum was divided by the number of questions in that case. TSM was divided by 11 because 11 questions made up the domain of social media activities. After getting the average for each of the TSM we have ATSM (Table 8) while for TAA we have ATAA (Table 9). Then the descriptive analyses were run for each to get the mean. The mean was used as the cut off in order to obtain the categorized average total social media (CatATSM) and which was divided into two groups. Those who had values of the mean and above were considered as people with high social media activities and those with scores below the mean were of low social media activities. For the categorized average academic activities scores was dived into two. Those who had scores below the mean were grouped to have poor academic activities while those with scores of the mean and above were considered to have good academic activities.

*Table 8: Descriptive Statistics for ATSM*

| Variable | ATSM | Valid N (listwise) |
|---|---|---|
| N Statistic | 396 | 396 |
| Range | 2.6364 |  |
| Minimum | 1.7273 |  |
| Maximum | 4.3636 |  |
| Mean | 3.099633 |  |
| Std. Deviation | 0.4671073 |  |
| Variance | 0.218 |  |
| Skewness (Statistic) | 0.137 |  |
| (Std. Error) | 0.123 |  |
| Kurtosis (Statistic) | -0.046 |  |
| (Std.Error) | 0.245 |  |

Mean of 3.099633 was used as cut off.

*Table 9: Descriptive Statistics for ATAA*

| Variable | ATAA | Valid N(listwise) |
|---|---|---|
| N Statistic | 396 | 396 |
| Range | 2.8333 |  |
| Minimum | 2.0000 |  |
| Maximum | 4.8333 |  |
| Mean | 3.318803 |  |
| Std. Deviation | 0.4827478 |  |
| Variance | 0.233 |  |
| Skewness (Statistic) | 0.168 |  |
| (Std. Error) | 0.123 |  |
| Kurtosis (Statistic) | 0.429 |  |
| (Std. Error) | 0.245 |  |

Mean of 3.318803 was used as cut off.

To find out whether there is any effect of social media activities on academic activities, binary regression was run, taking the CatATAA as the dependent variable while the CAtATSM was the independent variable. The level of statistical significance was set at $p \leq 0.05$ as shown in Table 10.

*Table 10: Variables in the Equation*

| Variable | Statistic |  |
|---|---|---|
| B | -0.522 | 0.495 |
| S. E | 0.207 | 0.159 |
| Wald | 6.359 | 9.746 |
| Df | 1 | 1 |
| Exp(B) | 0.594 | 1.641 |
| 95% C.I for Ep (B) Lower | 0.396 |  |
| Upper | 0.890 |  |

*1) Interpretation of result*

The reference category is low social media activity (Table 15). The column called EXP (B) shows the odds ratio for those who have low level of social media activities. This means that those who have high social media activities are less likely to have good academic activities level. There is a negative association by looking at the negative sign before the value in column labelled B. Therefore, the lower the level of social media activity the higher the likelihood of having a high level of academic activities.

The column called Sig, shows the p-value which is 0.012 which means that the association between social media activities and academic activities is statistically significant at *p=0.012*. The part of the table called 95% C.I. for Exp (B) gives us the lower confidence limits (lower) and the upper confidence limits (upper). The interval does not include the null value (which is always *x=1* in logistic regression) and, thus, the results are statistically significant. In summary, the odds ratio for good academic activities with high social media activity compared to those with low social media activity is 0.594 (95% CI 0.396, 0.890). This is significant, p-value of 0.012. We therefore accept the alternative hypothesis and reject the null. This agrees with [7] who found that while social media use for academic purposes does not show significant predictor for academic performance, using social media for nonacademic activities significantly negatively predicted academic performance and [13] affirmed that students who multitask during lectures have lower academic grades.

## 2. H1: Social media posts affect students' mood (get them angry or depressed.)

This cannot be tested using logistic regression because of insufficient data for analysis. We chose to apply independent t-test to the continuous data on mood after reading social media posts and categorical data of Average Total Social media score (ATSM).

*Table 11: Freq/percentage distribution of mood after social media posts*

| Variable | Freq. | Percent | Valid Percent | Cumulative Percent |
|---|---|---|---|---|
| Strongly Disagree | 47 | 11.9 | 11.9 | 11.9 |
| Disagree | 99 | 25.0 | 25.0 | 36.9 |
| Neutral | 148 | 37.4 | 37.4 | 74.2 |
| Agree | 83 | 21.0 | 21.0 | 95.2 |
| Strongly agree | 19 | 4.8 | 4.8 | 100 |
| Total | 396 | 100 | 100 | |

From Table 11, a total of 294 respondents out 0f 396 that participated in the study believe that social media posts do not affect their mood. These are participants that answered strongly disagree, disagree, or remained neutral (47,99 and 148) to the question on whether social media posts affect their mood (get the angry or depressed). 294 represents 74.3% of total participants. 102 (25.8%) accept that they are negatively affected by social media posts. These are participant that selected 'agree' or 'strongly agree' to the question on social media effect on bad mood.

*Table 12: Levene's Test for Equality of variance*

| Variable | | Statistic |
|---|---|---|
| Levene's Test for Equality of Variances | F | .382 |
| | Sig | .537 |
| t-test for Equality of Means | t | 2.613 |
| | df | 394 |
| | Sig. (2-tailed) | 0.009 |
| | Mean Difference | 0.276 |
| | Std. Error Difference | 0.106 |
| | 95% Confidence Interval of the difference | Lower | -0.483 |
| | | Upper | 0.068 |

### 2) Levene's Test for Equality of variance
The output in Table 12 shows that p-value is greater than 0.05 (p>0.05). p-value of 0.537 shows that there is equality of variance hence the Assumption of homogeneity of variance for application of t-test has been met. We have a t-statistic that is negative. This means that there is a reversal of directionality of the effect under study. This indicates that sample average value is bigger or smaller from the other to which it is compared to. Secondly, there is reduction in t-statistic and a large reduction in degrees of freedom, this has the effect of increasing *p*-value farther above the critical significance level of 0.05. We therefore reject the alternative hypothesis and accept the null hypothesis. The negation of t-statistic which gives a reversal of directionality of effects under study: "social media comments affect participant's mood (gets him angry or depressed), but that the participant's mood affects his social media comment. We therefore submit that social media posts are not a significant predictor for mood (angry, sad or depressed).

## 3. H1: Students struggle with time management as a result of frequent involvement in social media.

To test for this, first the scores that make up the total social media involvement were added up (TSMI). In order to get the scores to their original form, the sum was divided by 2 because questions made up the domain of social media activities. After getting the average for the TSMI we have ATSMI. Then the descriptive analyses were run to get the mean. The mean was used as the cut off in order to obtain the categorized average total social media involvement (CatATSMI) with those who had values of the mean and above were considered as people with high social media involvement and those with scores below the mean were of low social media activities.

To find out whether there is any relationship between of social media involvement and time management, binary regression was run, taking the CatATSMI as the dependent variable while the hours spent per day on social media was the independent variable. The level of statistical significance was set at p ≤ 0.05. See Table 13 for details.

*Table 13: Descriptive Statistics for Categorical Average Total Social Media Involvement*

| Variable | ATSMI | Valid N (listwise) |
|---|---|---|
| N Statistic | 396 | 396 |
| Range | 4,00 | |
| Minimum | 1.00 | |
| Maximum | 5.00 | |
| Mean | 3.1818 | |
| Std. Deviation | 0.83728 | |
| Skewness (Statistic) | -0.103 | |
| (Std. Error) | 0.123 | |
| Kurtosis (Statistic) | -0.282 | |
| (Std. Error) | 0.245 | |

For categorization of Average Total Social media involvement (CATSMI), score of the mean (3.1818) and above. High social media involvement and those with scores below the mean are of low social media involvement using a cut value of 0.500 as shown in Table 14.

*Table 14: Classification Table[a,b]*

|  | Observed | Predicted | | |
|---|---|---|---|---|
|  |  | Categorized Average Time Social Media Involvement | | Percentage Correct |
|  |  | Low Involvement | High involvement |  |
| Step 0 | Categorized Average Time Social Media Involvement — Low Involvement | 219 | 0 | 100.0 |
|  | High involvement | 171 | 0 | .0 |
|  | Overall Percentage |  |  | 56.2 |

a. Constant is included in the model. b. The cut value is 0.500

*Table 15: Variables in the Equation*

| Hrs/day | B | S.E | Wald | df | Sig. | Exp(B) | 95% C.I. for Exp(B) | |
|---|---|---|---|---|---|---|---|---|
|  |  |  |  |  |  |  | Lower | Upper |
| 13+ |  |  | 50.826 | 3 | 0.000 |  |  |  |
| 0-3 | -3.358 | 0.637 | 27.774 | 1 | 0.000 | 0.035 | 0.010 | 0.121 |
| 4-8 | -2.024 | 0.632 | 10.270 | 1 | 0.001 | 0.132 | 0.038 | 0.456 |
| 9-12 | -1.872 | 0.675 | 7.698 | 1 | 0.006 | 0.154 | 0.041 | 0.577 |
| Constant | 2.159 | 0.610 | 12.543 | 1 | 0.000 | 8.667 |  |  |

### 3) Interpretation of the result

The reference category for this analysis was those who spend 13hrs and above daily on social media. There is a negative association by looking at the negative sign before the value in column labelled B. Therefore, those who spend 0-3hrs per day on social media are less likely to have high social media involvement compared to those who spend 13hrs and above. Those in the 4-8hours also have a negative association (-2.024) and are therefore less likely to have high social media involvement. However, they have a higher likelihood of high social media involvement than those who spend 0-3hours. Similarly, those who spend 9-12hours on social media daily have a negative association (-1.872) also but with higher likelihood than the two previously mentioned. The OR is 0.035 (95% CI 0.010-0.121; $p < 0.001$).

We state that from the result, there is statistical significance, and the longer the hours spent on social media, the more students battle with time management as a result of high social media involvement. We therefore accept the alternative hypothesis and reject the null.

### V. CONCLUSION AND RECOMMENDATION

Our research purposes were two-fold: Firstly, to determine the effects of social media on students' academic activities, mood, and time management abilities from a Nigerian perspective, and secondly to find how the discovered effects relate to the students' academic performance. The result was presented in the result and discussion section. We found that association between social media and academic activities is statistically significant, however, a negative association exists between them which implies that the high the level of social media activity, the lower academic activities participation. Similar association was observed on the effects of social media on students' time management ability. Statistical tests showed there is statistical significance, but also with a negative association which implies that increase in social media activities result in decrease in ability to effectively manage time. Result on the effect of social media on students' mood (angry, depressed) showed no statistical significance and a negative t-value, indicating a reversal on directionality of effects sought. This means that social media posts do not result in bad mood, but higher the level of bad mood leads to higher social media posts.

We recommend that Nigerian institutions management bodies adopt the use of Signal Jammers for times of the day when academic activities are at its peak and in examination halls with the aim of controlling students multi-tasking social media with lectures, duration of social media use, as well as involvement/addiction level.

### VI. LIMITATION

Our second goal was to find the relationship of the sought effects on students' academic performance. This goal could not be achieved as the study was carried out at a time Nigerian academic staff union engaged in strike action and the students could not easily access the results for completed sessions.

### VII. FUTURE WORK

We shall advance the work by finding the relationship between the impacts of social media as revealed by our research, and students' academic performance.